\documentclass[prl,a4paper,superscriptaddress]{revtex4}
\usepackage{comment}
\usepackage{color}
\usepackage{amsmath}
\usepackage{amssymb}
\usepackage{graphicx}
\usepackage{braket} 
\usepackage{esint}
\usepackage{adjustbox} 
\usepackage[utf8]{inputenc}

\usepackage[unicode=true,pdfusetitle,
bookmarks=true,bookmarksnumbered=false,bookmarksopen=false,
breaklinks=true,pdfborder={0 0 1},backref=false,colorlinks=true]
{hyperref}
\hypersetup{
	linkcolor=red,urlcolor=blue,citecolor=blue,anchorcolor=blue} 
\usepackage{ulem} 

\usepackage{dcolumn}\usepackage{bm}

\makeatother

\begin{document}

\title{$PT$-Symmetric potential  impact on the scattering of a Bose-Einstein condensate from a Gaussian Obstacle}

\author{Jameel Hussain }
\affiliation{Department of Electronics, Quaid-i-Azam University, Islamabad, 45320 Pakistan}

\author{Muhammad Nouman}
\affiliation{Department of Physics, COMSATS University Islamabad, Pakistan}
\author{Farhan Saif} 
\affiliation{Department of Electronics, Quaid-i-Azam University, Islamabad, 45320 Pakistan}

\author{Javed Akram  }

\email{javedakram@daad-alumni.de}

\affiliation{Department of Physics, COMSATS University Islamabad, Pakistan}
 

\begin{abstract}
 The scattering of a Bose-Einstein Condensate (BEC) from a Gaussian well and Gaussian barrier is investigated over a wide range of depths and heights, respectively.  We compare analytical and numerical results for a BEC scattering from Gaussian Obstacles, both in the presence and in the absence of $PT$-symmetric potential.  And we find out that the Complex Ginzburg-Landau Equation (CGLE)  method has limitations due to the limited number of variational parameters of the ansatz. We also find that the presence of the $PT$-symmetric potential controls the reflection and the transmission flux of the BEC through the Gaussian Obstacle. 
\end{abstract}
 
 \date{\today}
 
\maketitle

\section{Introduction} \label{Sec-1}
The theoretical  and experimental realization of Bose-Einstein Condensate (BEC) created  many new possibilities to observe and analyze the quantum phenomena on macroscopic level \cite{clark1984einstein,Anderson-1995,PhysRevLett.75.3969}. Moreover, the confinement of the BEC in different  potential traps systems provide us better control over BEC \cite{PhysRevA.60.4171}, for instance, to study interference \cite{PhysRevA.72.021604}, solitons creation \cite{PhysRevA.81.053618,PhysRevA.93.023606,Akram_2016}, scattering \cite{PhysRevA.75.065602,PhysRevA.96.013618,Hussain2019}  tunneling \cite{PhysRevA.98.053629} and interaction of impurities with the BEC \cite{PhysRevA.93.033610,Akram_2018}.
\par
In 1998 Bender and Boettcher  presented the idea that some non-Hermitian Hamiltonians can have real spectrum \cite{PhysRevLett.80.5243}. Such Hamiltonians, with complex potentials follow the $PT$-symmetry, mathematically ensured by the condition $V(x)=V^{*}(-x)$, where $V(x)$ represents the external potential of the system. 
Bender and Boettcher idea is an extension to quantum mechanics from the real to the complex domain. $PT$-symmetric extension of quantum mechanics for non-hermitian Hamiltonian helps to encapsulate the idea of loss and gain in the system  \cite{Bender-2016,El-Ganainy:07,Ruschhaupt_2005}. The $PT$-symmetry has been manipulated and realized experimentally  in optics \cite{Christodoulides-2003,El-Ganainy:07}, and later on it is extended for the BEC \cite{PhysRevLett.101.150408,PhysRevA.82.013629}. 
\par  
In this paper, we consider the BEC scattering from the  Gaussian well  and Gaussian barrier in the absence of (conservative system), and in the presence  of (non-conservative) $PT$-symmetric environment.  For such a non-conservative $PT$-Symmetric system, in a double-well potential well, the atoms can be injected from one side and removed from the other side simultaneously   \cite{PhysRevLett.101.080402},   the injection and removal of atoms can be done by using laser radiation  \cite{PhysRevLett.110.035302,Gericke-2008},   the atom can be loaded at the desired side of the double well potential to ensure the exact compensation by atomic lasers \cite{Cirac_1996,Robins-2013,PhysRevA.87.051601}. We study analytically and numerically quasi-one-dimensional (1D) scattering dynamics of a BEC, the scattering includes both transmission and reflection from a Gaussian barrier. 
We study that the transmission and reflection can be controlled by the barrier height $V_0$. The $PT$-symmetric potential introduce another parameter to control the scattering characteristics of a BEC at the Gaussian barrier in the harmonic potential. In this respect, we organize this paper as follows. We discuss the theoretical model in Sec.~\ref{Sec-2}. A comparison of analytical and numerical results is given in Sec.~\ref{Sec-3}. In Sec.~\ref{Sec-4}, we present the scattering of a BEC from a repulsive barrier in the absence of $PT$-symmetric potential. Later, in Sec.~\ref{Sec-5}, we study the impact of $PT$-symmetric potential on the scattering of a BEC from a repulsive Gaussian barrier. The summary and conclusion is discussed in Sec.~\ref{Sec-6} and  the last Sec.~\ref{Sec-7} is assigned for acknowledgment.

\section{Theoretical Model}  \label{Sec-2}
The BEC dynamics in a quasi-1D regime is governed by the quasi-1D GPE, \cite{PhysRevLett.81.938,Petrov-2004,Gross-1963,citeulike:7001053} 

\begin{eqnarray}
\iota\hbar\frac{\partial\psi(x,t)}{\partial t}=\left[-\frac{\hbar^{2}}{2m}\frac{\partial^{2}}{\partial x^{2}}+V(x)+g_{s}|\psi(x,t)|^{2}\right]\psi(x,t),
 \label{eq1}
\end{eqnarray} 
with the normalization condition $\int|\psi(x,t)|^{2}dx=1$. Here  $\psi(x,t)$ describes the wavefunction of the condensate, $m$ represents mass of individual atom,  $t$ defines time and the $x$ stands for a 1D-space coordinate. The BEC experiences external potential $V(x)=V_h+V_g+W(x)$, with $V_h$ denotes the harmonic potential, $V_g$ illustrates the Gaussian potential barrier, and $W(x)$ stands for the complex $PT$-symmetric potential. 
The interaction strength can be described as $g_{s}=2N\hbar\omega_{r}a_{s}$,  where $a_{s}$ represents the s-wave scattering length \cite{PhysRevA.93.033610}, $N$ stands for the number of atoms  in the BEC and $\omega_{r}$ characterizes the radial frequency component of the trap \cite{PhysRevA.93.023606}. To do numerical simulation, we make 1D-GPE (\ref{eq1}) dimensionless. Therefore, we measure time in $\omega_{x}^{-1}$, length of the harmonic oscillator along x-axis in $\sqrt{\hbar/{m\omega_{x}}}$ and energy in $\hbar \omega_{x}$. The quasi-1D GPE reduce to the dimensionless form,
\begin{eqnarray}
\iota\frac{\partial\psi(x,t)}{\partial t}=\left[-\frac{1}{2}\frac{\partial^{2}}{\partial x^{2}}+V(x)+g_{s}|\psi|^{2}\right]\psi(x,t),
\label{eq2}
\end{eqnarray}

 with dimensionless interaction strength reduce to  $g_{s}=2N\omega_{r}a_{s}/(\omega_{x}L)$, here $L=\sqrt{\hbar/m\omega_{x}}$ denotes the length of the oscillator along x-axis and the dimensionless external potential is given by, 
 \begin{equation}
  V=\frac{x^{2}}{2}+V_{0}.e^{-x^2}+i.W_{0}.x.e^{-x^2},
 \end{equation}
 here, $V_0$ defines the dimensionless Gaussian well depth ($V_0<0$) and Gaussian barrier height ($V_{0}>0$). While, the gain ($W_0>0$) and the  loss ($W_0<0$) of the BEC atoms can be controlled by the strength of $PT$-symmetric potential  $W_{0}$. In this paper, we do not study the impact of the width of the Gaussian obstacle on the dynamics of the BEC, therefore we chosen a constant width ``1''. However, according to our understanding, the width controls the tunneling between the adjacent wells, which needs a detailed investigation.   
 

\subsection{Analytical method} \label{Sec-2a}
The variational approach gives the analytical information about the numerical solution of the system  
 \cite{PhysRevA.27.3135,PhysRevA.56.1424}.  Here in this paper, we investigate the analytical method to compare and validate our numerical simulations results. Additionally, we want to explore the limitations of the analytical technique.  The analytical method relies on the choice of the initial normalized trial wave-function. 
Here, we let the initial  normalized ansatz as 

\begin{eqnarray}
\psi(x,t)= \frac{1}{\sqrt{a(t)\sqrt{\pi}}} e^{-\frac{(x-x_0(t))^2}{2 a(t)^2}+i x   \alpha(t)+i x^2\beta(t) }, 
\label{eq3}
\end{eqnarray} 
where $x_0(t)$ represents the mean position of the BEC. Here $a(t)$ defines the dimensionless width of the BEC, $\alpha(t)$  represents the velocity of the BEC and $\beta(t)$  represents the time-dependent derivative of the width of the BEC. To study our system analytically, we take the Lagrangian density as, 
\begin{eqnarray}
\mathcal{L} =   \frac{i}{2}\left(\psi \frac{\partial \psi^*}{\partial t} -
\psi^* \frac{\partial \psi}{\partial t} \right) - \frac{1}{2}|\frac{\partial\psi }{\partial x}|^2 +  
 V(x)|\psi|^2+\frac{g_s}{2} | \psi|^4. \label{eq4}
\end{eqnarray}
By using above Lagrangian density, we get the Lagrangian of the system  $L=\int\mathcal{L}dx.$ We begin by writing the total Lagrangian of the system as a sum of two terms, i.e., $L=L_c+L_{nc}$ where $L_c$ represents the conservative  term  and $L_{nc}$ describes the non-conservative part of the Lagrangian. Here, the conservative system means the potential without complex part of the external potential  while  non-conservative term represents the complex part of the external potential. By using the above Langrangian, we determine the complex Ginzburg-Landau equation (CGLE) as \cite{refId0,Ankiewicz-2007,PhysRevE.92.022914,Hu-2017},
\begin{eqnarray}
\frac{d}{dt}\Bigg(\frac{ \partial L_c}{\partial \dot{q}}\Bigg)-\frac{ \partial  L_c}{ \partial q}=2Re\Bigg[\int\limits_{-\infty}^\infty  W(x)\psi \frac{\partial \psi^\ast}{\partial q}dx\Bigg], \label{eq5} 
\end{eqnarray}
where $q$ stands for variational parameters $a(t)$, $x_0(t)$, $\alpha(t)$ and $\beta(t)$.
By using Eq.~(\ref{eq4}) and Eq.~(\ref{eq5}), we  calculate the  time-dependent equation for the mean position of the BEC, 
\begin{eqnarray}
x_0''(t)+x_0(t)=\frac{2V_0x_0(t)}{\xi(t)^\frac{3}{2}} e^{\frac{-x_0^2(t)}{\xi (t)} } \label{eq6},
\end{eqnarray}
where $\xi(t)=1+a^2(t)$,  here, we ignore the non-conservative term, however, we give the detailed non-conservative equation in appendix A.  
The dimensionless width of the BEC changes with time as it collides with the Gaussian barrier. To understand this phenomenon, we determine analytically the dimensionless time-dependent equation for the width of the BEC as,
\begin{eqnarray}
a''(t)+a(t)=\frac{1}{a^3(t)}+ \frac{g_s}{\sqrt{2\pi}a^2(t)}+  
\frac{2V_0a(t)}{\xi(t)^\frac{3}{2}} e^{-x_0^2(t)/\xi(t)}  \left[1
 -\frac{2 x_0^2(t)}{\xi(t)} \right],
\label{eq7} 
\end{eqnarray}
while deriving above Eq.~(\ref{eq7}) again we neglect the complex part of the potential, $L_{nc}$.  "Non-conservative" $L_{nc}$ part makes our equations cumbersome, therefore that equation is presented in appendix A.

\subsection{Numerical method} \label{Sec-2b}
To numerically simulate our research problem, we perform discretization of the dimensionless quasi-1D GPE Eq.~(\ref{eq2}). We take space-step as $\bigtriangleup x=0.0177$ and we choose the time step as $\bigtriangleup t=0.0001$. Here, we use the time-splitting spectral method  
\cite{BAO2003318,Vudragovic12,Kumar15,Loncar15,Sataric16,Zhu_2016}. To get the numerical equilibrium results for the shifted harmonic potential $V=(x-35)^2/2$, we use strange-split method, where the ground state wavefunction for different interaction strength is achieved by simulating in imaginary-time  $\tau=\iota t$.  
For the dynamical evolution of the wavefunction of the BEC, it is worthwhile to mention that the ground state wavefunction serves as an initial condition for the rest of the numerical simulations.

\section{A comparison of Analytical and Numerical Results} \label{Sec-3}
To study the limitation of analytical results as discussed in Sec.~\ref{Sec-2a}, we compare them with numerical results produced in Sec.~\ref{Sec-2b}. The comparison of analytical and numerical results are presented in Fig.~(\ref{Fig1}-\ref{Fig4}) for attractive Gaussian well and  for a repulsive Gaussian barrier. In order to understand the basic scattering behavior, we choose the dimensionless interaction strength as $g_s=30$. To compare analytical and numerical results, for attractive Gaussian well, we plot the temporal density graph in the absence of $PT$-symmetry in Fig.~\ref{Fig1}(a-b) and in Fig.~\ref{Fig2}(a-b) for a dimensionless Gaussian well depth $V_0=-500$, and $V_0=-1000$ respectively.  Similarly, in  Fig.~\ref{Fig1}(c-d) and in  Fig.~\ref{Fig2}(c-d), we present the numerical and analytical results for the scattering of a BEC in the presence of $PT$-symmetry environment for the same dimensionless interaction strength and  well depths as mentioned above. 
Initially, we place a BEC in this potential $V={(x-35)^2}/{2}$. Later, the BEC set into motion by quenching the trapping potential minima to coincide with the maxima of the Gaussian barrier as $V=x^2/2+V_0 e^{-x^2}$.  To analyze our results, we divide our research problem into three different subsections, for attractive Gaussian well,  low repulsive Gaussian barrier and   high repulsive Gaussian barrier.
\subsection{Attractive Gaussian well}
\begin{figure}
\centering
\includegraphics[height=8cm,width=9cm]{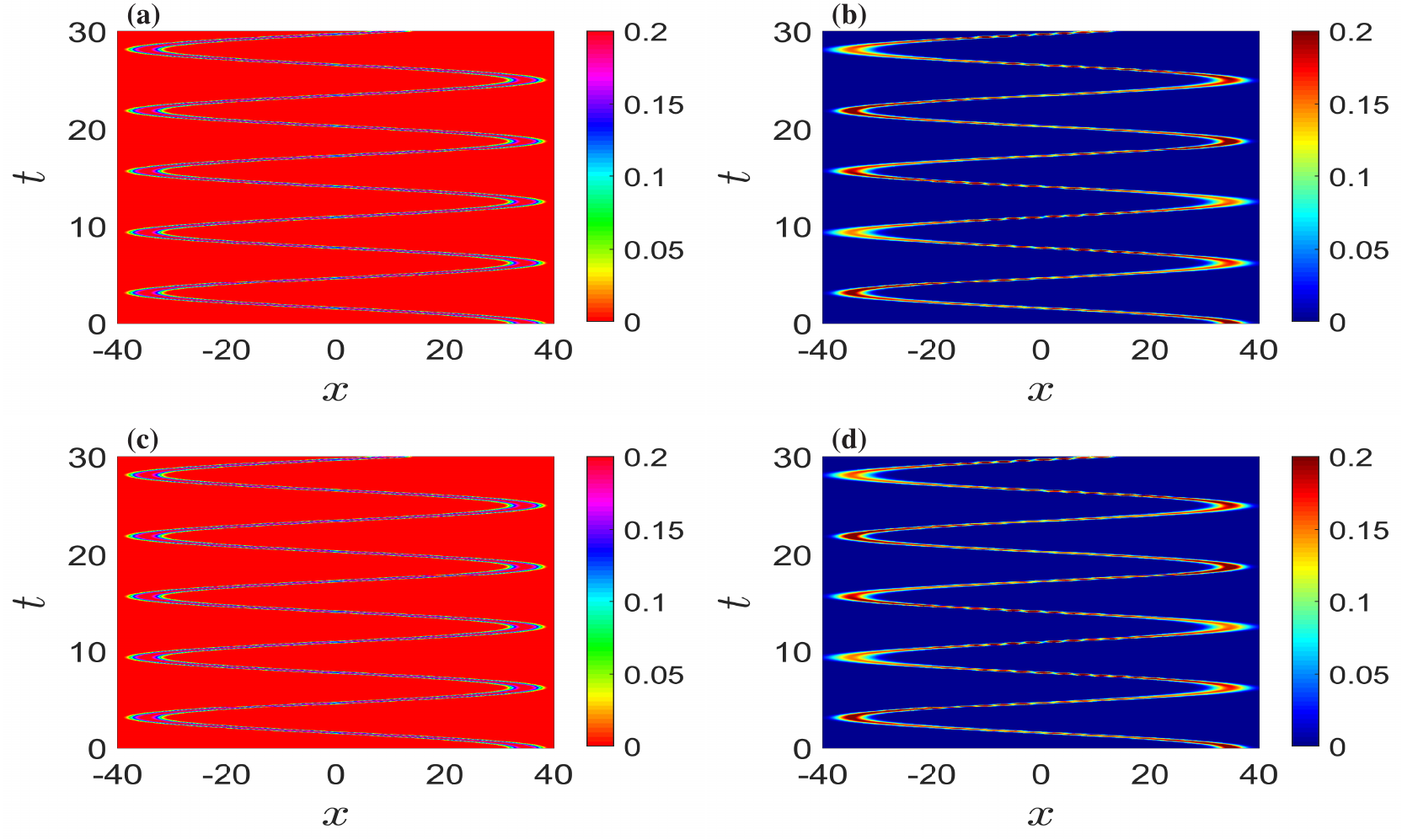}  
\caption{(Color online) Numerical (left column) and analytical (middle column) density profile of the BEC scattering from an attractive Gaussian well $V_0=-500$. The   dimensionless parameters are, $g_s=30$, and $x_0=35$. The Fig. (a,b) are the cases without $PT$-symmetric potential $W_0=0$, while for Fig. (c,d) the $PT$-symmetric potential is  $W_0=1$. While the (right column) describes the time dependence of the mean-position and the width of the BEC wave-function. }
\label{Fig1}
\end{figure}

\begin{figure}
\centering
\includegraphics[height=8cm,width=9cm]{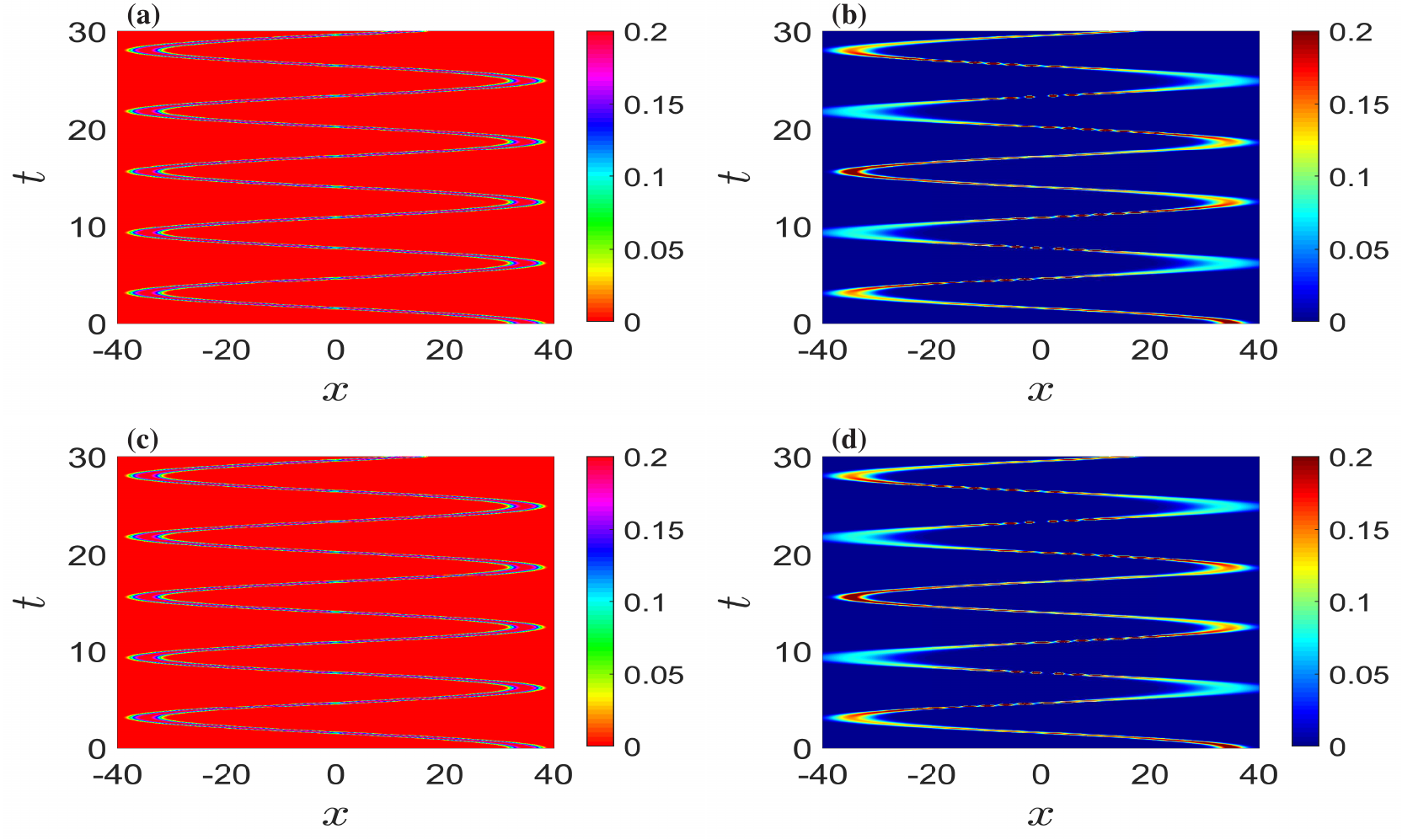}  
\caption{(Color online) Comparison of numerical (left column) and analytical (right column) results for the scattering of a BEC from an attractive Gaussian well $V_0=-1000$. The dimensionless parameters are, interaction strength $g_s=30$, BEC is initially placed at $x_0=35$. The Fig. (a) and (b) are the cases without $PT$-symmetric potential. While for Fig (c) and (d) the dimensionless strength of the $PT$-symmetric potential is $W_0=1$.}
\label{Fig2}
\end{figure}

The BEC initially placed at $x_0=35$, later at time $t=0$ the BEC get a kick due to quenching of the potential and it starts moving towards the attractive Gaussian well. In Fig.~\ref{Fig1} and Fig.~\ref{Fig2}, we can see that, for an attractive Gaussian well, $V_0=-500$ and $V_0=-1000$, respectively, in the absence of $PT-$symmetric environment, the analytical and numerical results match over a wide range of Gaussian well depths. In Fig.~\ref{Fig1}, we find out that the numerical and analytical results agree in the absence  of $PT-$symmetric potential,  Fig.~\ref{Fig1}(a-b), i.e., $W_0=0$,  and in the presence  of $PT-$symmetric potential,  Fig.~\ref{Fig1}(c-d),  i.e., $W_0=1$,  very well with each other for $V_0=-500$. Later, we plot Fig.~\ref{Fig2} for a very high attractive Gaussian well, $V_0=-1000$, where analytical and numerical results are matches in the absence Fig.~\ref{Fig2}(a-b), i.e., $W_0=0$ and mismatches in the presence of $PT$-symmetry Fig.~\ref{Fig2}(c-d), i.e., $W_0=1$. We note that  the complex Ginzburg-Landau Eq.~\ref{eq5} valid for a large range of attractive Gaussian barriers well in the absence of $PT$-symmetric environment. However, CGLE Eq.~\ref{eq5} could not capture the physics of scattering of a BEC from an attractive Gaussian well in the presence of $PT$-symmetry for $W_0\geq1$. 

\subsection{Low Repulsive Gaussian Barrier}
\begin{figure}
\centering
\includegraphics[height=8cm,width=9cm]{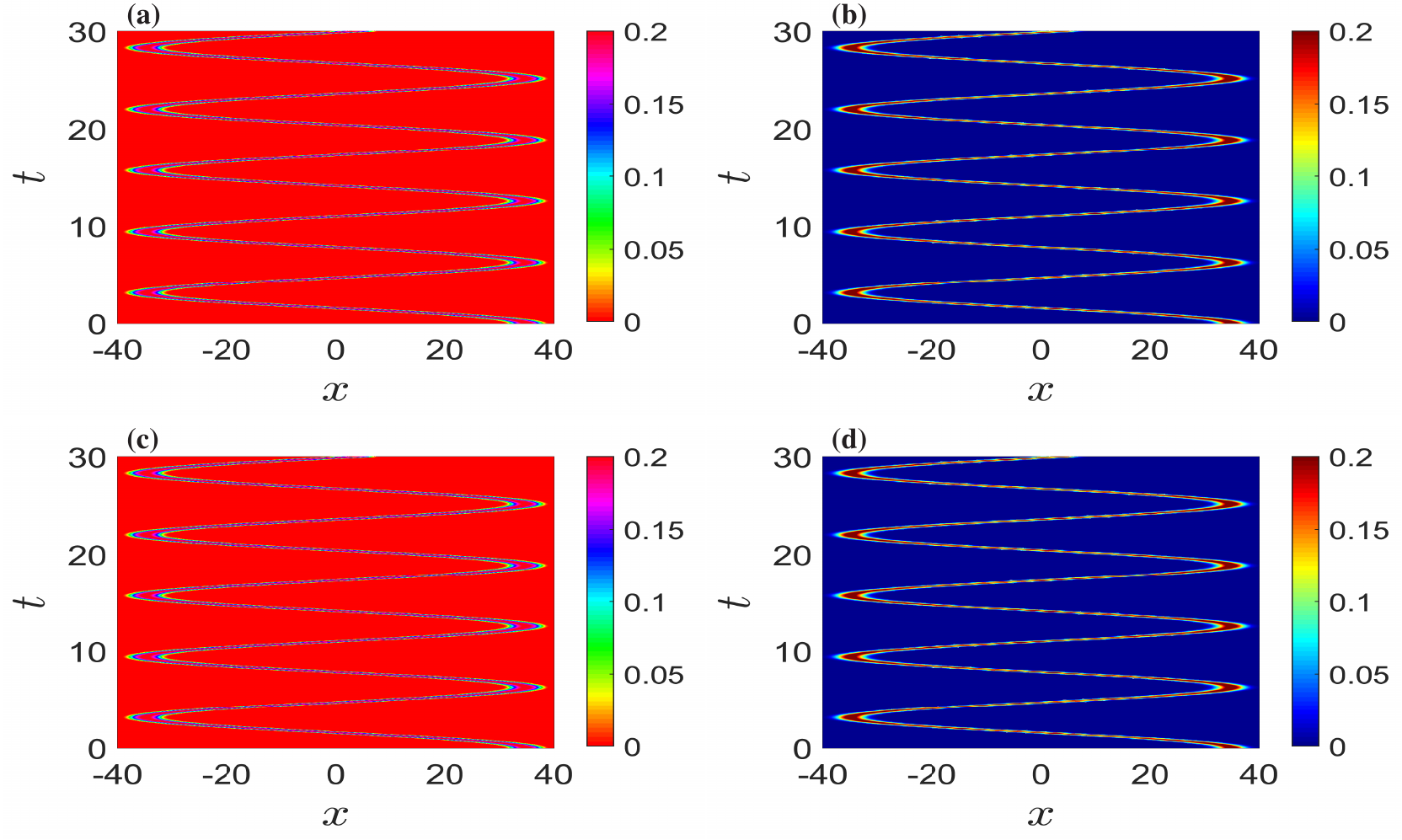}  
\caption{(Color online) Comparison of numerical (left column) and analytical (right column) results for the scattering of a BEC  from a  low repulsive Gaussian barrier $V_0=100$. The dimensionless parameters are, interaction strength $g_s=30$, BEC  is initially placed at $x_0=35$. The Fig. (a) and (b) are the cases without $PT$-symmetric potential. While for Fig. (c) and (d) the dimensionless strength of the $PT$-symmetric potential is $W_0=0.1$.}
\label{Fig3}
\end{figure}

For a  low repulsive barrier i.e., $V_0=100$, and in the absence of $PT$-symmetric environment, we compare analytical and numerical results in   Fig.~\ref{Fig3}(a-b). While in Fig.~\ref{Fig3}(c-d), we  show results in the presence of $PT$-symmetric environment both numerical and analytical. We note in Fig.~\ref {Fig3}(a,b) for a  low barrier height, both the numerical and analytical results agree with each other. We find no discrepancy for the temporal density plot in Fig.~\ref{Fig3}(c-d), where the complex potential strength is $W_0=0.1$.  Therefore, we can safely conclude that for a   low  repulsive Gaussian barrier, the complex-Ginsburg-Landau Eq.~(\ref{eq5}) capture the physics of scattering of a BEC,   for a weak $PT$-symmetric potential.  

\subsection{High Repulsive Gaussian Barrier}
\begin{figure}
\centering
\includegraphics[height=8cm,width=12cm]{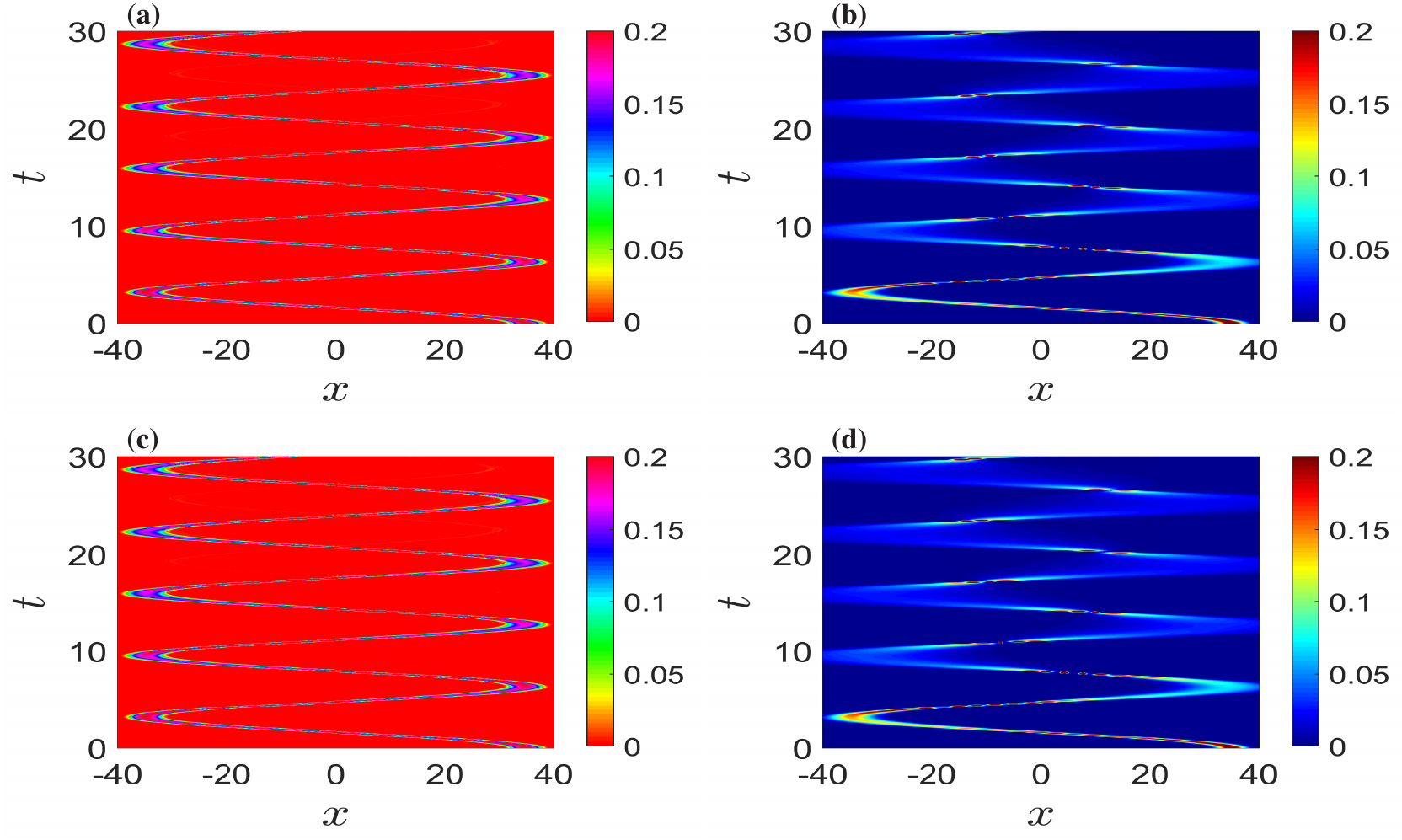} 
\caption{(Color online) Comparison of numerical (left column) and analytical (right column) results for the scattering of a BEC  from a large repulsive Gaussian barrier $V_0=500$. The dimensionless parameters are, interaction strength $g_s=30$, BEC is initially placed at $x_0=35$. The Fig. (a) and (b) are the cases without $PT$-symmetric potential. While for Fig. (c) and (d) the dimensionless strength of the $PT$-symmetric potential is $W_0=0.1$.}
\label{Fig4}
\end{figure}

In this subsection, we compare the  analytical and numerical results for the  scattering of a BEC from a high repulsive Gaussian barrier $V_0=500$. The analytic and numerical results are shown in Fig.~\ref{Fig4}(a,b) and  Fig.~\ref{Fig4}(c,d) in the absence and in the presence of $PT$-symmetry potential environment respectively. We realize that the analytical results  mismatch with the numerical simulations, as presented in Fig.~\ref{Fig4}.
The reason for such a mismatch and the limitations of analytical results lies in the choice of the ansatz in Eq.~(\ref{eq3}).  
The initial ansatz, a Gaussian, deforms its shape during the collision with repulsive Gaussian barrier. We notice that  higher the barrier, larger the deformation and hence larger is the mismatch. To get a good analytical result one needs a good ansatz with large ensemble of variational parameters. Such a large ensemble can make the problem cumbersome and notoriously complicated. Moreover, the collision of a BEC with the barrier generates the quasi-particles  on the surface of the  BEC, which makes it more hard for analytical analysis. Therefore, for the high repulsive Gaussian barrier more than $V_0=500$, the analytic results appears to lose the credibility. In our case, analytic results have limitations, it loses the true picture behind the physics of scattering of the BEC from the repulsive Gaussian barrier. Thus, from now on for the rest of this paper, we will rely only on our numerical simulations.  

\section{Numerical Results In The Absence of $PT$-Symmetric Potential} \label{Sec-4}
\begin{figure}	
\centering
	\includegraphics[height=5.5cm,width=12cm]{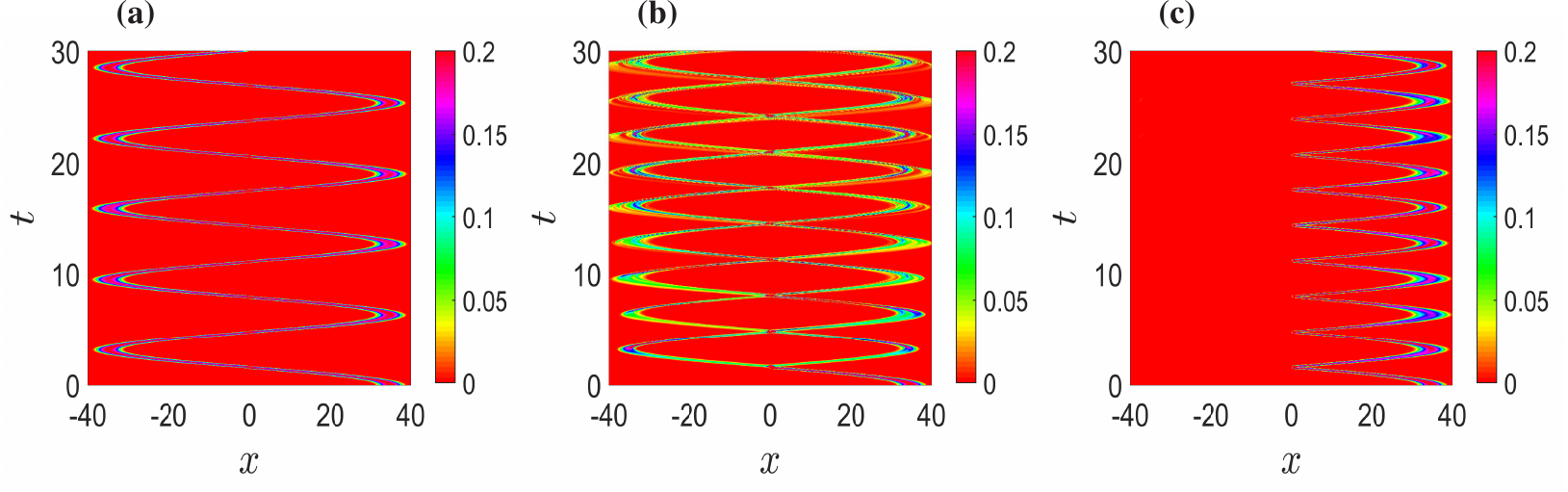}  
	\caption{(Color online) Numerically obtained temporal density graph shows the scattering of a BEC, in the absence of $PT$-Symmetric potential with dimensionless parameters $g_s=30$, and $x_0=35$, from different repulsive barrier heights (a) $V_0=400$, (b) $V_0=600$, and (c) $V_0=700$.  }
	\label{Fig5}
\end{figure}

In this section, we discuss numerical results of a BEC scattering from a mild, high  and very-high repulsive barrier for a conservative system, i.e., in the absence of $PT$-symmetric environment as shown in  Fig.~\ref{Fig5}.
\begin{figure}
\centering
	\includegraphics[height=9cm,width=12cm]{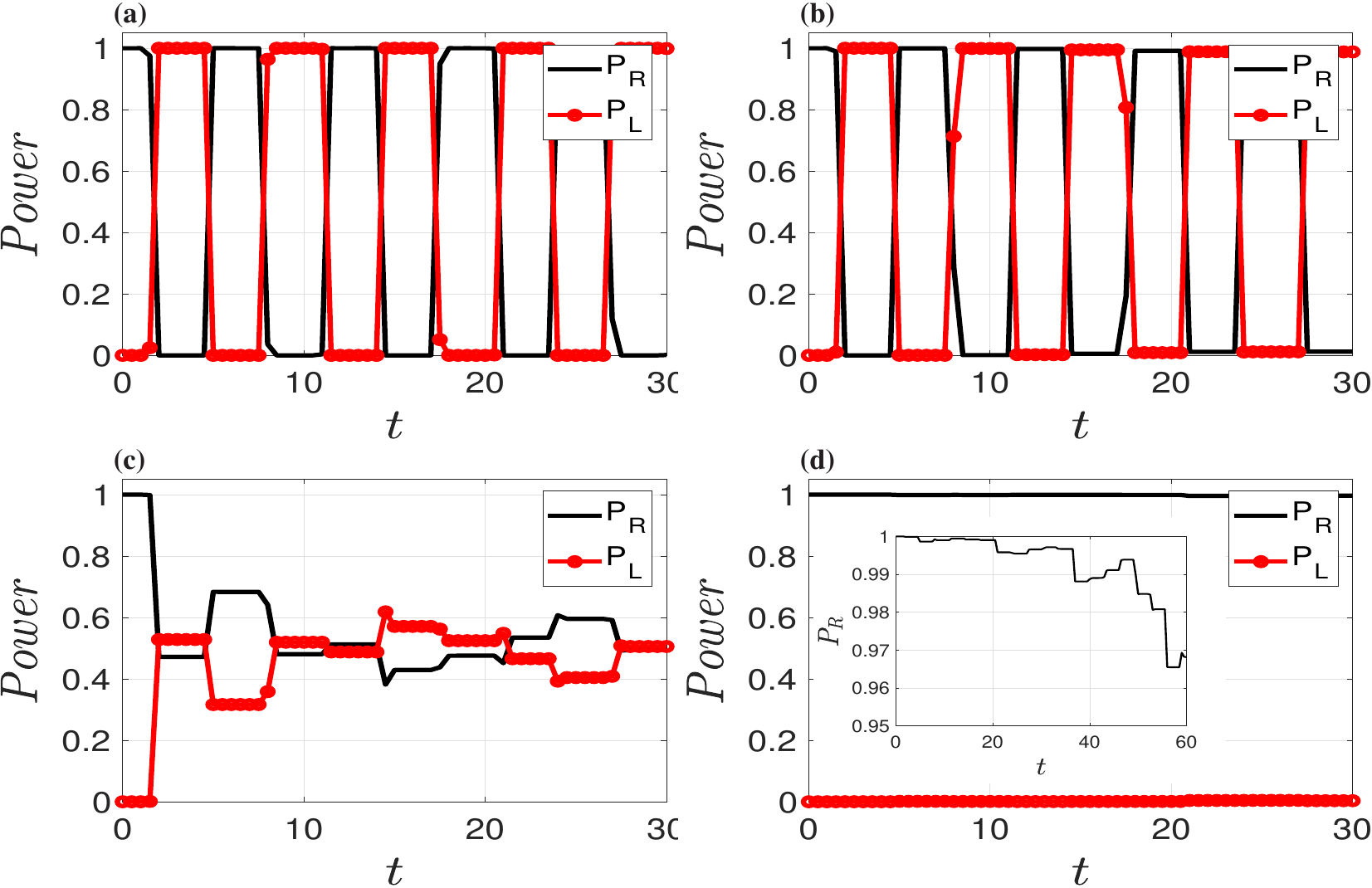}
	\caption{(Color online) Power of a BEC versus the dimensionless time. The dimensionless parameters are $x_0=35$, $g_s=30$ and Gaussian barrier heights are (a) $V_0=400$, (b) $V_0=500$, (c) $V_0=600$, and (d) $V_0=700$.  The inset showing the decrease of a right side power ($P_R$) of the BEC with respect to time.}
	\label{Fig6}
\end{figure} 
In Fig.~\ref{Fig5}(a) and Fig.~\ref{Fig4}(a), we plot the temporal density graph of the BEC for  the barrier heights  $V_0=400$ and $V_0=500$, respectively. For these barrier heights the BEC
unable to see the  low barriers at all.  The BEC does not see these barrier heights. Apparently the BEC experiences a harmonic confinement and performs to-and-fro motion.  In Fig.~\ref{Fig5}(c) for high Gaussian potential barrier $V_0=600$, the BEC exhibit scattering, here, we observe reflection and transmission of the BEC at the barrier. In Fig.~\ref{Fig5}(d) for a very-high barrier height $V_0=700$ the BEC  shows     total reflection.   Hence,  the BEC is  confined to the same side of the  double well  potential, where it was placed initially. This happen due to the very-high barrier height, however, one can observe a small tunneling of the BEC for a comparable large time. To quantify, this information, we plot the power of the BEC for the \textit{Right} and the \textit{Left} side of the Gaussian barrier. Here the \textit{Left} side power (LSP) is defined as $P_L=\int_{-\infty}^{0} |\psi |^2dx$ and the \textit{Right} side power (RSP) is represented as $P_R=\int_{0}^{\infty}|\psi |^2dx$. Both powers RSP and LSP, fluctuate between ``1'' and ``0". We see from Fig.~\ref{Fig6}(a) that the initially BEC start  moving towards the barrier  from right side of the external potential.  Therefore,  initially the $P_R$ starts from its maximum value "1" and $P_L$ begins from "0". However, as BEC moves in time, the power of the wavefunction start oscillating between $P_L$ and $P_R$ as shown in  Fig.~\ref{Fig6}(a). Nevertheless, as we increase the barrier height from $V_0=500$ to $V_0=600$, we observe that the $P_L$ and $P_R$ starts oscillating. We also observe that both  $P_L$ and $P_R$  tends to  converge to $0.5$, which shows presence of transmission and reflection of the BEC from the barrier, as shown in Fig.~\ref{Fig6}(c) and BEC is divided into two fragments as already depicted in Fig.~\ref{Fig5}(c). For $V_0=700$, the \textit{Right} side power, $P_R$, of the BEC remains towards the \textit{Right} side of the external potential, where it was trapped  earlier.  That is a  clear sign of the confinement of a BEC on one side of the external Gaussian barrier potential. However, we also observe for a large dimensionless time, $t>20$, a small amount of BEC tunnel through the barrier, which is a quantum mechanical effect, it can be made more visible by increasing the simulation time as shown in subset of Fig.~\ref{Fig6}(d). 

\section{Numerical results In The Presence of $PT$-Symmetric Potential} \label{Sec-5}
\begin{figure}
\centering
	\includegraphics[height=5.5cm,width=12cm]{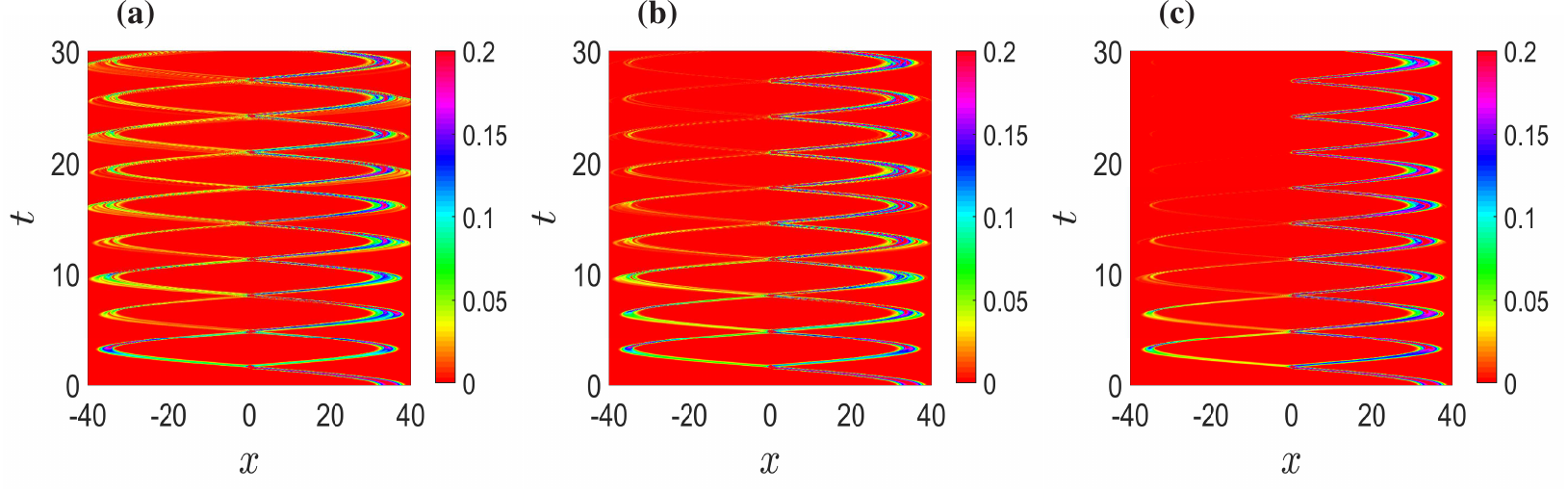}  
	\caption{(Color online) Numerically obtained temporal density graph shows the scattering of a BEC, in the presence of $PT$-Symmetric potential (a) $W_0=1 $, (b)  $W_0= 5 $, and (c) $W_0=  10$ and other parameters are  $V_0=600$, $g_s=30$, and $x_0=35$.}
	\label{Fig7}
\end{figure}

\begin{figure}
\centering
	\includegraphics[height=9cm,width=12cm]{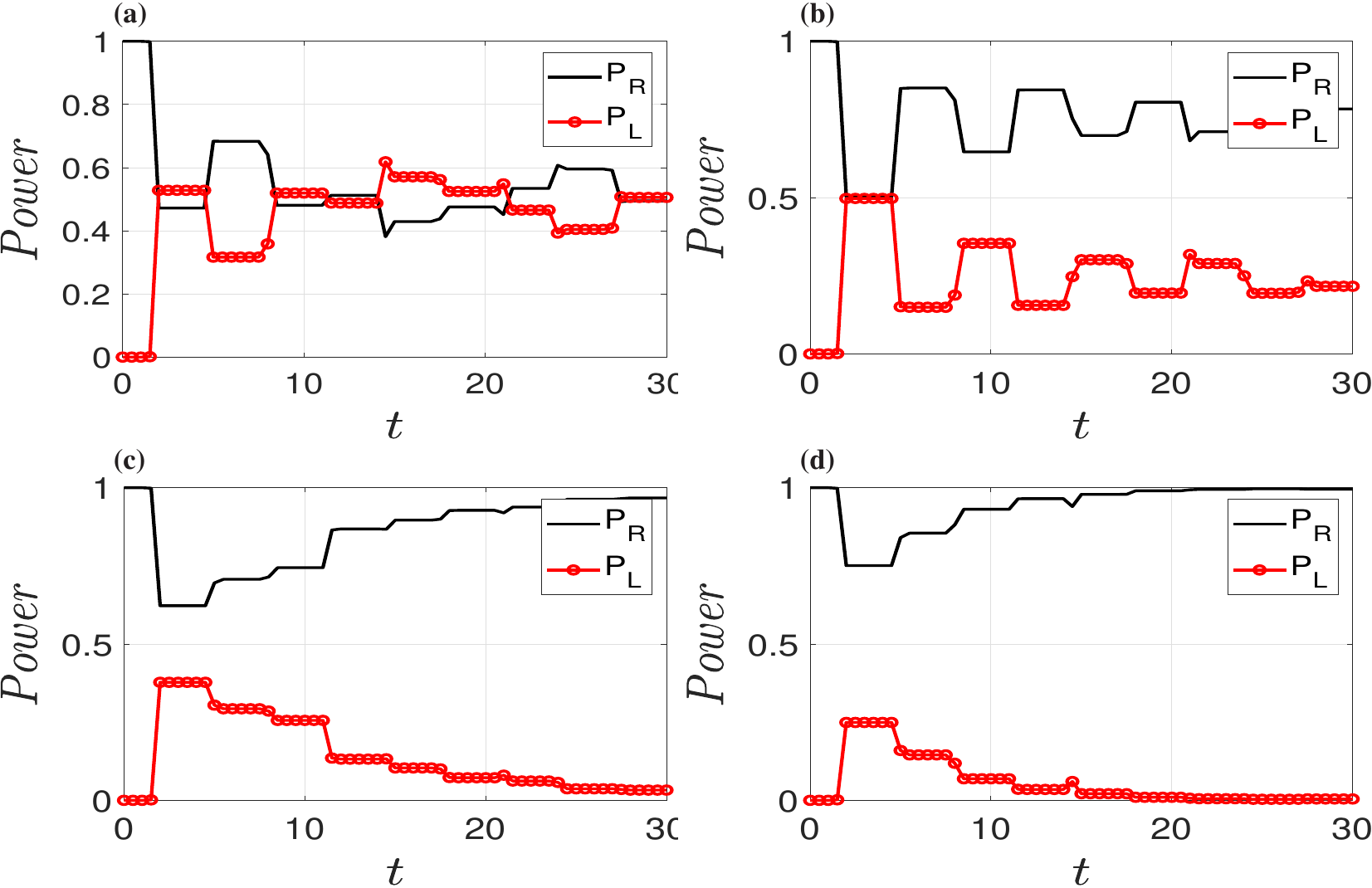}  
	\caption{(Color online) Power of a BEC under a $PT$-symmetric environment.     The amount of  dimensionless $PT$-symmetry potentials are (a) $W_0=0$, (b) $W_0=1 $, (c) $W_0= 5 $ and (d) $W_0=  10$.  Other dimensionless parameter are $x_0=35$,  $g_s=30$  and $V_0=600$.} 
	\label{Fig8}
\end{figure}

In this section, we discuss the numerical results of scattering of a BEC from a repulsive Gaussian barrier under a $PT$-symmetric environment as shown in Fig.~\ref{Fig7}. The Fig.~\ref{Fig7}(a-c) shows that the scattering of a BEC has been greatly influenced by the presence of loss and gain. Here the right side ($x>0$) of the $PT$-symmetric  potential represents the gain and the left side ($x<0$) describes the loss in the system.  Fig.~\ref{Fig5}(b) shows that the BEC is divided into two fragments for a barrier height of $V_0=600$ for $W_0=0$. The Fig.~\ref{Fig7}(a-c)  for $W_0=1 $, $W_0= 5 $ and $W_0=  10$  reveals that the presence of loss and gain environment changes the reflection and transmission amount of the BEC  considerably. Scattering of the BEC can be made more visible by plotting the \textit{Left} side power, $P_L$ and \textit{Right} side power, $P_R$ of the BEC as shown in Fig.~\ref{Fig8}(a-d). Here, we find out that as we increase the $PT$-Symmetric potential "$W_0$", the $P_R$ starts growing with time. For example, for a specific case, Fig.~\ref{Fig8}(c) for $W_0=5$, we note that initially the $P_R$ starts decreasing due to the BEC's multiple  reflections from the  Gaussian barrier but gradually the amount of $P_R$ rises as time goes on. Thus the $PT$-symmetric potential influence the transmission and tunneling of a BEC.
We also observe in Fig.~\ref{Fig8}(d), that by  increasing the amount of $PT$-symmetric potential $W_0=10$, the BEC stops penetrating into the left side of the potential from very early time as compared with the same potential barrier height. We also note that as time goes on the transmission of the BEC halt. Hence the $P_R$ gradually grows to ''1`` and $P_L$ decreases to ``0''. 
We can safely say that the presence of the $PT$-symmetric potential turns  the system into a unidirectional medium. The term ``unidirectional''  means that, we can control the direction of the  transmission and tunneling of a BEC through a Gaussian barrier in the presence of $PT$-symmetric potential. 
Hence tuning to a specific reflection or transmission coefficient is possible through the external $PT$-symmetric potential.
By using our technique, it also seems possible that one can calibrate the system for a specific amount of transmission/reflection for a desire time by just controlling the $PT$-symmetric potential.

\section{Conclusion} \label{Sec-6}
In this paper, we study the impact of the presence of  a $PT$-symmetric potential on the BEC transmission and reflections from a Gaussian well and Gaussian barrier. We find that scattering from the Gaussian barrier
can be controlled by a combination of the barrier height and a $PT$-symmetric potential. In the first part of this research paper, we compare numerical simulation results with the analytical results obtained by using the Complex Ginzburg-Landau equation. We conclude that the CGLE captures the physics of the scattering of a BEC from a Gaussian well and and from a  low Gaussian barrier. However, the analytical results fail  to capture the physics of the BEC scattering if a small amount of $PT$-symmetric, $W_0>0.1$, is present.  We observe that the transmission and reflection at the barrier can be tuned by choosing a proper amount of $PT$-symmetry potential. In transistors, the "base" terminal controls the current through emitter and collector. Here, in our case, such $PT$-symmetric potentials control  the transmission of the BEC in such a way that it behaves like a transistor, where the complex part of the potential serves like a "base". Moreover, our proposed model can be used as an atomic beam-splitter, in our proposed study by controlling the height of the barrier we can divide the density of the atomic beam into two parts for the varying density size. Additionally, by controlling the $PT$-symmetric strength, we can steer atomic beam-splitter density with respect to time, which can be seen as an additional feature of an atomic beam-splitter.
 \par 
\par 

\section{Acknowledgment} \label{Sec-7}
Jameel Hussain gratefully acknowledges support from the COMSATS University Islamabad for providing him a workspace.

\section{Appendix A}

A complete solution for the  Mean position of the BEC including ''non-conservative`` part of the Lagrangian

\begin{eqnarray} 
 x_0''+x_0 =&  \frac{2 V_0 x_0 e^{-\frac{x_0^2}{\zeta }}}{\zeta ^{3/2}}+ \frac{W_0^2 x_0 \left(a^4+a^2+2 x_0^2\right) \left(2 a^6+a^4+\left(4 a^2+2\right) x_0^2-a^2\right) e^{-\frac{2 x_0^2}{\zeta }}}{a^2 \zeta ^6} \nonumber \\
&  + \frac{W_0 e^{-\frac{x_0^2}{\zeta }}}{a \zeta ^{9/2}} \Big[2 a \zeta  x_0 \left(-a^4+a^2-2 x_0^2+2\right) x_0'   \nonumber \\
&    +a' \left(3 a^2 \zeta ^2+4 a^2 x_0^4+2 \left(a^4-4 a^2+1\right) \zeta  x_0^2\right)\Big]  
\end{eqnarray}
 
Here, to write above equation into compact form we used these definitions  $x_{0}  \equiv x_{0} (t)$, $a \equiv a(t)$. 
Width of the BEC including  ''non-conservative``  part  
\begin{eqnarray} 
&& \frac{2 W_0^2 x_0^2 \left(2 a^6+a^4+\left(4 a^2+2\right) x_0^2-a^2\right){}^2 e^{-\frac{x_0^2}{\zeta }}}{a \sqrt{\zeta }}+ \nonumber \\
&& +2 a \zeta  W_0 \left(a x_0 a' \left(\left(8 a^2+4\right) x_0^4+\left(-2 a^4+11 a^2-2\right) \zeta ^2+2 \left(2 a^4-11 a^2-3\right) \zeta  x_0^2\right) \right. \nonumber  \\
&& \left. +\zeta  \left(2 a^8+3 a^6-4 \left(2 a^2+1\right) x_0^4-a^2+\left(-4 a^6+10 a^4+20 a^2+6\right) x_0^2\right) x_0'\right) \nonumber \\
&& - 4 a^3 \zeta ^4 V_0 \left(\zeta -2 x_0^2\right) = - 2 a^2 \zeta ^{13/2} e^{\frac{x_0^2}{\zeta }} \left(-\frac{1}{a^3}-\frac{g_s}{\sqrt{2 \pi } a^2}+a''+a\right)  \label{eq9}
\end{eqnarray} 
 
\textbf{We solve these two coupled differential equations by using Mathematica command $NDSolve$. $NDSolve$ commonly solves differential equations by using the Implicit-Runge-Kutta method or Explicit-Runge-Kutta method, depending on the type of equations. Indeed, both coupled differential equations are harder to solve, therefore, it is recommended to solve GPE numerically. }

\section{References} 

\end{document}